\documentclass[Review,sagev,times]{sagej}

\usepackage{moreverb,url}

\usepackage[colorlinks,bookmarksopen,bookmarksnumbered,citecolor=red,urlcolor=red]{hyperref}

\newcommand\BibTeX{{\rmfamily B\kern-.05em \textsc{i\kern-.025em b}\kern-.08em
T\kern-.1667em\lower.7ex\hbox{E}\kern-.125emX}}

\def\volumeyear{2016}

\begin{document}

\runninghead{Sherina, McCall, and Love}

\title{Fully Bayesian imputation model for non-random missing data in qPCR}

\author{Valeriia Sherina\affilnum{1}, Matthew N. McCall\affilnum{1,2} and Tanzy M. T. Love\affilnum{1}}

\affiliation{\affilnum{1}Department of Biostatistics and Computational Biology, University of Rochester Medical Center, USA\\
\affilnum{2}Department of Biomedical Genetics, University of Rochester Medical Center, USA}

\corrauth{Tanzy M. T. Love, Department of Biostatistics and Computational Biology,
University of Rochester Medical Center,
Rochester, NY,
USA.}

\email{tanzy\_love@urmc.rochester.edu}

\begin{abstract}
We propose a new statistical approach to obtain differential gene expression of non-detects in quantitative real-time PCR (qPCR) experiments through Bayesian hierarchical modeling. We propose to treat non-detects as non-random missing data, model the missing data mechanism, and use this model to impute Ct values or obtain direct estimates of relevant model parameters. A typical laboratory does not have the resources to perform experiments with a large number of replicates; therefore, we propose an approach that does not rely on large sample theory. We aim to demonstrate the possibilities that exist for analyzing qPCR data in the presence of non-random missingness through the use of Bayesian estimation. Bayesian analysis typically allows for smaller data sets to be analyzed without losing power while retaining precision. The heart of Bayesian estimation is that everything that is known about a parameter before observing the data (the prior) is combined with the information from the data itself (the likelihood), resulting in updated knowledge about the parameter (the posterior). In this work we introduce and describe our hierarchical model and chosen prior distributions, assess the model sensitivity to the choice of prior, perform convergence diagnostics for the Markov Chain Monte Carlo, and present the results of a real data application.
\end{abstract}


\maketitle

\section{Introduction}
There are three missing data mechanisms defined in the statistical literature: missing completely at random (MCAR), missing at random (MAR), and missing not at random (MNAR)~\citep{rubin1976inference, little2014statistical}. Our group previously showed that the data in qPCR experiments are most likely MNAR~\citep{McCall}. Even when the proportion of missing values in a qPCR dataset is not substantial, it is not appropriate to ignore or remove the genes or samples with missing values~\citep{rubin1976inference, little2014statistical}. In both likelihood-based inference and Bayesian, if the parameters describing the measurement process are functionally independent of the parameters of the missing data mechanism, the missingness process is called ignorable, and a non-random process is non-ignorable~\citep{molenberghs2005models}. In MNAR the parameters of the missingness process and the parameters of the measurement process are not functionally independent, and hence the process is non-ignorable. This means that in MNAR data we need to explicitly address the missing data mechanism, because ignoring it can lead to invalid inference about the model parameters.

To properly characterize the missing data mechanism in qPCR data, we first need to have a good understanding of the experimental procedures and the related properties of the measurement instrument. In a typical qPCR experiment, we start with samples mixed with primers and nucleotides for the identification and multiplication of the biological material. During each cycle we expect a doubling of the existing transcript. Due to the dynamic, real time nature of the qPCR experiment, many factors can contribute to the presence of missing values in the final qPCR data. For example, low abundance genes may fail to amplify or the qPCR instrument may have trouble detecting a weak signal from low-abundance genes even if amplification occurs. Therefore, the smaller the signal, the higher the probability of observing a non-detect, a reaction that failed to be detected by the qPCR technology. Even though these experiments involve a detection threshold, instead of a censoring model, we utilize a probabilistic model proposed by~\cite{McCall}, and model the missing data process as a function of the transcript value to be measured.

In this manuscript, we propose a new approach to obtain differential gene expression for all genes, including non-detects in qPCR experiments through Bayesian hierarchical modeling. 
Bayesian analysis does not assume large samples, as is the case with Maximum Likelihood Estimation (MLE); typically smaller data sets can be analyzed without losing power while retaining precision~\citep{hox2012few, hamra2013markov}. 
For more technical details we kindly refer the reader to~\cite{gelman2014bayesian}. 

The heart of Bayesian estimation is that everything that is known about a parameter before observing the data (the prior) is combined with the information from the data itself (the likelihood), resulting in updated knowledge about the parameter (the posterior). The prior information can stem from a meta-analysis, previous studies with comparable research populations, a pilot study, experts, or a range of other sources. If such knowledge is used we call the prior informative, and if no knowledge is available (or used) we call the prior uninformative. Estimation of Bayesian models is frequently done through Markov Chains Monte Carlo (MCMC). Unlike deterministic Maximum Likelihood algorithms, MCMC is a stochastic procedure that repeatedly generates random samples that characterize the posterior distribution of the parameters of interest. This is distinct from the sampling distribution of estimators which are estimated by MLE. The process of generating the random samples is the role of the Markov Chain. The process of generating summary statistics from those random samples is the role of Monte Carlo integration.

In the study of~\cite{galindo2004bayesian}, the authors specified an informative prior and concluded that the more information that is captured by well-specified priors the smaller the parameter bias. We consider a situation in which, a priori, there is some information available about the parameters of interest, so we use uninformative priors, weakly informative priors, or priors reflecting partial-knowledge of the parameters.
In summary, our main objective is to demonstrate methods to analyze small sample size data through the use of Bayesian estimation, and to determine the conditions under which Bayesian inference out performs MLE. This will guide the analysis of qPCR data in the presence of MNAR data.

We first introduce and describe our hierarchical model and chosen prior distributions and perform a prior sensitivity analysis. We then present the results of a real data application, assess the convergence of the Markov Chain Monte Carlo, compare parameter estimates with existing methods described in~\cite{Sherina231621}, and comment on open problems. 

\section{Modeling} \label{sec:ch2model}
Throughout, we analyze the gene expression of $I$ genes. We define a gene or feature by $i$. Each gene is measured under different conditions, perturbations, and in multiple replicates, we will refer to these conditions and perturbations as $k$ sample-types throughout the paper. $J_{(k)}$ denotes a partition of the samples into K sets of replicates. \\
    $Y_{ij}$ is the observed gene expression of a gene $i$, sample $j$ in $J_{(k)}$ - a sample of sample type $k$ ( $i = 1, \cdots, I; \, j = 1, \cdots, J; \, k = 1, \cdots, K$). For each $Y_{ij}$ there is a $Z_{ij}$ such that if $Z_{ij}=1$, the gene expression is observed, and if $Z_{ij}=0$, the value is a non-detect, with
	\begin{eqnarray}
	Pr(Z_{ij}=1) = \begin{cases} g(Y_{ij}) & \mbox{if } Y_{ij} < 40 \\
	0 & \mbox{if otherwise}, \end{cases}
	\label{eq:prob}	
	\end{eqnarray}
here, $g(Y_{ij})$ is a function that models the dependence between the gene expression and the probability of the value being observed. The following logistic regression model is a natural choice of such a relationship: 
	\begin{eqnarray}
	logit\Big(Pr(Z_{ij}=1) \Big)=\beta_0 + \beta_1 \mu_{ij}, 
	\label{eq:logit1}	
	\end{eqnarray}
where $\beta_0$ and $\beta_1$ are the logistic regression coefficients, and $\mu_{ij}$ is an estimate of a gene expression for gene $i$, sample $j$ that is potentially unobserved. One of the challenges that arise with the use of $\mu_{ij}$ in Equation~\ref{eq:logit1} is the presence of non-detects. When we do not have enough information to generate estimates for the individual $\mu_{ij}$, a possible solution is to borrow information across replicates in $J_k$. 

This is the data generating model proposed in~\cite{McCall} and
described in more details in~\cite{Sherina231621}. In this paper, we consider the $absolute$ gene expression, and model it as follows:
	\begin{eqnarray}
	Y_{ij} = \begin{cases} \theta_{ik} I(j \in J_k) + \delta_{j} + \varepsilon_{ij} & \mbox{if } Z_{ij}=1 \\
    \mbox{non-detect} & \mbox{if } Z_{ij}=0, \end{cases} 
	\label{eq:mod11}	
	\end{eqnarray}
where $Y_{ij}$ is again the observed gene expression or the $\Delta$Ct value from qPCR experiment, $\theta_{ik}$ is the true gene expression for gene $i$ in the sample-type $J_k$, $\delta_{j}$ represents a global shift in expression across samples, $\varepsilon_{ij}$ captures the stochastic component of the model.

Bayesian statistics combines the knowledge from the data in the likelihood with prior information. The prior distribution should reflect what we know about the parameters in the model. The main parameters of interest are the average differential gene expression and its variability (denoted by $\theta$ and $\sigma^2$) together with the missing mechanism parameters, $\beta_0$ and $\beta_1$. 

All the parameters in the model need to have a prior distribution. The specification of the prior distributions consists of three steps. First, background knowledge is needed as input for the specification. Such knowledge can stem from a meta-analysis, previous studies with comparable research populations, a pilot study, experts, or a range of other sources. In our case, we used existing knowledge of the range of parameters as inspiration. Second, for all the parameters a type of distribution has to be specified. We used a Normal distribution, denoted by $N(\cdot,\cdot)$, for $\theta$'s; for the standard deviation parameters we specified a Uniform distribution, denoted by $Unif(0,A)$; and for the probability of $Z=1$ we used a Bernoulli distribution, denoted by $Bern(\cdot)$. As~\cite{gelman2006prior} pointed out, inferences are not sensitive to the choice of $A$ for at least 3 groups and sufficiently large finite $A$.

The third step of the prior specification is to determine the shape of the prior distribution by means of choosing values of the hyper-parameters. For the mean parameters $\theta$, the hyper-parameters are the mean and the standard deviation of the normal distribution, denoted by $\theta_0$ and $\gamma$. We fixed the hyper-parameter for the mean of the average gene expression at $\theta_0 = 60$. This constant is larger than the maximum possible observed $\Delta$Ct, so that it pulls all the values of non-detects to be somewhat greater than the maximum possible observed $\Delta$Ct value. However, in the situation where all the replicates of a sample are non-detects, we really do not have any information about what the values should have been. Therefore, an estimate of the $\Delta$Ct value being 60 or 80 tells us that there is no confidence in the presence of gene expression. 
For the standard deviation parameters $\sigma$, the hyper-parameters are the maximum of the uniform range, this prior is equivalent to an $inverse-\chi^2$ with -1 degrees of freedom, $p(\sigma^2) \propto \sigma^{-1}$~\citep{gelman2006prior}.
We model $\beta_0$ and $\beta_1$ jointly as $bivariate\, Gaussian$ with a zero mean vector, a variance of 100, and zero correlation. In addition, $\beta_1$ is restricted to be positive.

Below is a full specification of our Bayesian Model for the observed data ($Y_{ij}, Z_{ij}$):
\par
{Model:} 

\noindent $Y_{ij} \sim N(\mu_{ij}, \sigma^2_i) \times I(0,\infty),$ where
$\mu_{ij} = \theta_{ik} + \delta_{j}, \, j \in J_k,$ \par
\noindent $Z_{ij} \sim Bern(p_{ij})$, and $logit(p_{ij}) = \beta_0+\beta_1 \times \mu_{ij}.$  
\par
{Prior distributions for model parameters:}

\noindent $\theta_{ik} \sim N(\theta_0 = 60, \gamma^2_i)$, $\sigma_i \sim Unif(0, A),$ and ${\underline \beta} \sim N_2(\underline{\mu}_\beta,B) \times I_{\beta_1>0},$ where $A = 10 \text{ or } 100$, $\underline{\mu}_\beta=(0, 0)'$, and $B$ is $diag_2(100)$.

Prior for hyper-parameters $\gamma_i$:
\noindent $\gamma_i \sim Unif(0, A).$

\subsection{Posterior distributions} \label{sec:ch2poster}
In this section we derived the full conditional distributions of the parameters of interest and the hyper-parameter $\gamma^2$. Some of them do not have a known distributional form.
The distribution of the hyper-parameter $\gamma^2$, the variance of the mean parameter $\theta$, is:
\begin{eqnarray}
\gamma_i^2|rest & \sim & Scaled\, Inv-\chi^2 \Big(J-1, \sum_{j\in
J_{(k)}}(\theta_{ij}-\theta_0)^2 \Big), \text{ where } \theta_0 = 60. \nonumber
\end{eqnarray}

\begin{eqnarray}
p(\underline\beta|rest) & \propto & 				                                \exp(-\dfrac{1}{2} \underline\beta B^{-1} \underline\beta') \times I_{\beta_1>0} \nonumber \\
				& \times  & 
				\prod_{i=1}^I \prod_{j\in J_{(k)}} \Big( \big( \frac{\exp((1,\mu_{ij})\underline\beta)}{1+\exp((1,\mu_{ij})\underline\beta)} \big)^{ Z_{ij}} \nonumber \\
				& \times  & \big(\frac{1}{1+\exp((1,\mu_{ij})\underline\beta)} \big)^{1-Z_{ij}} \Big)
\end{eqnarray}

\begin{eqnarray}
\sigma^2_i|rest & \sim & Scaled\, Inv-\chi^2 \Big(J-1, \sum_{j\in
J_{(k)}}(Y_{ij}-\theta_{ij}-\delta_{j})^2 \Big)
\end{eqnarray}

\begin{eqnarray}
p(\theta_{ij}|rest) & \propto & \prod_{i=1}^I N \Big( \mu_\theta, \Sigma_\theta \Big) \nonumber \\
                &\times & \prod_{j\in J_{(k)}} \Big( \big( \frac{\exp((1,\mu_{ij})\underline\beta)}{1+\exp((1,\mu_{ij})\underline\beta)} \big)^{ Z_{ij}} \nonumber \\
				& \times  & \big(\frac{1}{1+\exp((1,\mu_{ij})\underline\beta)} \big)^{1-Z_{ij}} \Big)
\end{eqnarray}

where $$\mu_\theta = ((\sigma^2_{i})^{-1}+ (\gamma_i^2)^{-1})^{-1} \times ((\sigma^2_{i})^{-1}\frac{1}{J_{(k)}}\sum_{j \in J_{(k)}}(Y_{ij}-\delta_{j})+(\gamma_i^2)^{-1}\times\theta_0),$$ and $$\Sigma_\theta = ((\sigma^2_{i})^{-1}+ (\gamma_i^2)^{-1})^{-1}.$$

\subsection{Parameter estimation}
We consider two ways of summarizing the output from the MCMC:
\begin{enumerate}
    \item Fully Bayesian Imputation (FBI). By taking a median of all draws for the parameters of interest, $\theta$s and $\sigma^2$s, we acquire direct parameter estimates of the mean and the variance of gene expression. 
    \item Single Imputation Bayesian (SI Bayes). By taking the median of draws of missing $y$, we obtain single imputation for individual missing data points. We further calculate means and the variances of gene expressions to get the values of parameters for this method. Note that we expect the variance to be under-estimated here because we are ignoring the fact that some $y$ are missing and treating estimates of $y$ as data.
\end{enumerate}

\section{Simulation study}
Based on the design and the parameters of the experimental data from~\cite{sampson2013gene} we selected the true values of the model parameters to be used in our simulation study. We set $\beta_0=-35.7$ and $\beta_1=1$, the number of genes to 16, the number of replicates, the length of $J_{(k)}$, within each sample type to 6, and the number of sample types $K$, we also set to 6. We assume the missing mechanism is common across all genes.

To generate synthetic data, we first simulate complete data, then remove part of the data by constructing a MNAR data mechanism. In the first step, we generate three components: true $\theta_{ik}$, $\delta_j$, and  $\varepsilon_{ij}$. We generate $\theta_{ik}$ from $N(\mu_\theta, \sigma^2_{\theta} I)$, we set $\sigma^2_{\theta}=3$ and $\mu_{\theta}$ we generate from a  $N(31, 3.5^{1/2})$ truncated from 20 to 40.5. $\delta_j$ we set to 0, and simulate $\varepsilon_{ij}$ from $N(0, \sigma^2_{i} I)$, where true $\sigma^2_i$ are coming from $Unif(0.06, 1.3)$.
In the second step we generate the missing data indicators from $Binom(p_{ij})$, where $p_{ij}$ is the probability of a point being missing according to the logit model proposed in the Equation~\ref{eq:logit1} calculated for each data point. All the individual Ct values $\ge$ 40 and values with missing indicators are replaced with a value of 40, indicating a non-detect.

We compared performance under several prior distributions and assessed sensitivity to the choice of prior. For every scenario we ran 100 simulations. We summarize the results as the 25$^{th}$, 50$^{th}$, and 75$^{th}$ quantiles of bias and MSE for all genes and samples. We have 16 genes and 6 sample-types, so for each synthetic data set there are 16 distinct values of $\sigma^2_{i}$ and 16$\times$6=96 different $\theta_{ik}$. We show results for FBI because they are better, results for SI Bayes are shown in the application.

\subsection{Prior sensitivity analysis}
The importance of prior sensitivity analysis is hard to overestimate; the choice of prior distributions is influential on the estimation results. The more information that is captured by well-specified priors the smaller resulting parameter bias.
We looked at several choices of prior distribution for variability components of the model. We specified priors for $\sigma$ and $\gamma$, or $\sigma^2$ and $\gamma^2$. We compared the conjugate inverse gamma priors with varying parameters and weakly informative uniform priors with two different boundaries: (0, 10) and (0, 100). The summary of Bias and MSE for the mean ($\theta$) and the variance components ($\sigma^2$) are given in Table~\ref{tab: priorBiasMSE}.

\begin{table}[ht]
\centering														
\begin{tabular}{c|ccc|ccc}		
  \hline														
	&		&	Bias	&		&		&	MSE	&			\\
  \hline
  \multicolumn{7}{c}	{$\sigma^2 \sim$ Unif(0, 100), $\gamma^2 \sim$ Unif(0, 100)}			\\
  \hline														
  $\theta$	& $_{	-0.032	}$&	-0.011	& $_{	0.012	}$&$_{	0.100	}$&	0.122	& $_{	0.137	}$	\\
  $\sigma^2$	& $_{	0.025	}$&	0.033	& $_{	0.040	}$&$_{	0.033	}$&	0.048	& $_{	0.062 }$	\\
  \hline														
\multicolumn{7}{c}	{$\sigma^2 \sim$ Unif(0, 10), $\gamma^2 \sim$ Unif(0, 10)}			\\
  \hline														
  $\theta$	& $_{	-0.003	}$&	0.016	& $_{	0.047	}$&$_{	0.099	}$&	0.120	& $_{	0.138	}$	\\
  $\sigma^2$	& $_{	0.026	}$&	0.036	& $_{	0.044	}$&$_{	0.034	}$&	0.049	& $_{	0.064	}$	\\
  \hline														
 \multicolumn{7}{c}	{$\sigma^2 \sim$ IG(0.001, 0.5), $\gamma^2 \sim$ IG(0.1, 0.1)}			\\
  \hline														
  $\theta$	& $_{	-0.031	}$&	-0.010	& $_{	0.013	}$&$_{	0.102	}$&	0.124	& $_{	0.139	}$	\\
  $\sigma^2$	& $_{	0.042	}$&	0.047	& $_{	0.049	}$&$_{	0.034	}$&	0.048	& $_{	0.061	}$	\\
  \hline														
\multicolumn{7}{c}	{$\sigma^2 \sim$ IG(0.001, 0.5), $\gamma^2 \sim$  IG(0.001, 0.5)	}			\\
  \hline														
  $\theta$	& $_{	-0.031	}$&	-0.010	& $_{	0.013	}$&$_{	0.102	}$&	0.124	& $_{	0.139	}$	\\
  $\sigma^2$	& $_{	0.042	}$&	0.047	& $_{	0.049	}$&$_{	0.034	}$&	0.048	& $_{	0.061	}$	\\
  \hline														
\multicolumn{7}{c}	{$\sigma^2 \sim$ IG(0.001, 0.5), $\gamma^2 \sim$ IG(1, 10)	}			\\
  \hline														
  $\theta$	& $_{	-0.030	}$&	-0.008	& $_{	0.014	}$&$_{	0.103	}$&	0.124	& $_{	0.139	}$	\\
  $\sigma^2$	& $_{	0.042	}$&	0.046	& $_{	0.049	}$&$_{	0.034	}$&	0.048	& $_{	0.061	}$	\\
  \hline																		
\multicolumn{7}{c}	{	$\sigma^2 \sim$ IG(0.001, 0.5), $\gamma^2 \sim$ IG(0.001, 0.001)	}			\\
  \hline														
  $\theta$	& $_{	-0.031	}$&	-0.010	& $_{	0.013	}$&$_{	0.102	}$&	0.124	& $_{	0.139	}$	\\
  $\sigma^2$	& $_{	0.042	}$&	0.047	& $_{	0.049	}$&$_{	0.034	}$&	0.048	& $_{	0.061	}$	\\
  \hline														
\multicolumn{7}{c}	{ $\sigma^2 \sim$ IG(1, 10), $\gamma^2 \sim$ IG(0.001, 0.001)}			\\						
  \hline														
  $\theta$	& $_{	-0.025	}$&	-0.004	& $_{	0.016	}$&$_{	0.208	}$&	0.232	& $_{	0.247	}$	\\
  $\sigma^2$	& $_{	0.646	}$&	0.670	& $_{	0.701	}$&$_{	0.577	}$&	0.587	& $_{	0.621	}$	\\
  \hline														
\multicolumn{7}{c}	{$\sigma^2 \sim$ IG(0.001, 0.001), $\gamma^2 \sim$ IG(0.001, 0.001) }			\\						
  \hline														
  $\theta$	& $_{	-0.032	}$&	-0.011	& $_{	0.012	}$&$_{	0.096	}$&	0.118	& $_{	0.132	}$	\\
  $\sigma^2$	& $_{	0.003	}$&	0.008	& $_{	0.011	}$&$_{	0.029	}$&	0.041	& $_{	0.054	}$	\\
    \hline														
\end{tabular}
\caption{Varying prior distributions for $\gamma^2$ and $\sigma^2$. \label{tab: priorBiasMSE}}
\end{table}	

The $\theta$ estimates are unbiased, while $\sigma^2$ is slightly overestimated, meaning that this fully Bayesian model incorporates additional uncertainty into the estimation of the variance. MSE's are higher for $\theta$ and lower for $\sigma^2$ due to the bias variance trade off. The estimation results vary for $IG(w_1,\, w_2$) priors, in contrast, the results for $Unif(0,\,A)$ priors are consistent and almost identical across different ranges. Therefore, in the real data application, we decided to use the uniform distribution. If a researcher knows more information about the distribution of the parameters, they have the ability to specify the prior in our model.

\section{Experimental data application}
To study the performance of the proposed methods and compare them to existing methods of imputing MNAR data, we used an experimental study of the effect of p53 and/or Ras mutations on gene expression with 3 or 4 replicates per sample-type~\citep{mcmurray2008synergistic}. Real data applicability is an essential part of testing new methods, and when testing new methods for missing data, it is important to know the true values. 

The data set has a small amount of missing values, about 3\%. To have a complete data set, we started with removing all the genes with missing values. We call this data set \textit{complete}. Thereafter, we replaced all the $\Delta$Ct values greater or equal to 30 with 30, this resulted in a \textit{truncated} data set with 2\% of missing values found in 12\% of unique genes. We use this dataset in testing FBI and SI Bayes. We initiated MCMC and obtained 10000 draws from posterior distributions of the parameters $\beta$, $\theta, \sigma^2, \text{ and } y$. The average effective sample size of the parameters is 4872. We calculated estimates using FBI and SI Bayes methods. 
Thereafter, we calculate the difference between parameter estimates from experimental data and from several existing methods, such as a Penalized EM algorithm incorporating missing-data mechanism (PEMM), a Direct Estimation (DirEst), a Multiple Imputation (MI), and present the results in Figure~\ref{fig:1}.
We compared the proposed FBI and SI Bayes methods to estimates simply using the truncated data (Trunc), a penalized EM algorithm incorporating non-random missingness (PEMM) proposed by~\cite{chen2014penalized}, and MLE based methods: DirEst and MI, proposed by our group~\citep{Sherina231621}. PEMM calculates $Pr(Y|Z=1)=\text{constant}\times \exp^{(-\phi\times Y)}$ for positive $Y$, where $\phi$ is the parameter in the missing-data mechanism; we set $\phi=1$. We calculated $\hat \theta$ and $\hat \sigma^2$ for PEMM, DirEst, MI, SI Bayes, and FBI, and compared them to the estimates from the \textit{complete} data.

We then repeated the same process with truncating $\Delta$Ct values at 33. This time the resulting data has overall 1\% of missing values that occur in 7\% of unique genes. The average effective sample size in the MCMC draws for the parameters is 4980. The results of this study are presented in Figure~\ref{fig:2}.

\subsection{FBI outperforms MLE based methods}
When there is a small \% of missing data, there is very little information about the missing data mechanism, common across all the genes. Our Bayesian modeling framework allows for better estimation of the missing mechanism in the presence of a small amount of non-detects. There has been a lot of attention drawn to methods that can accommodate large proportions of missing data, some methods have shown to be successful, for example ~\cite{chen2014penalized}. In a laboratory setting with limited resources, researchers work with only a few replicates per condition, and missing data is rather unusual. When the missing data mechanism is non-ignorable, researchers are more likely to utilize and rely on methods that are tailored to estimate the missing data mechanism better with less available information about it.


We see that for the data truncated at 30, DirEst and MI slightly outperform fully Bayesian imputation in estimating the average gene expression, while SI Bayes substantially underestimates gene expression in comparison to PEMM and even truncated data mean estimates. The FBI shows very similar performance to DirEst and MI in terms of bias for $\sigma^2$, while SI Bayes has higher bias, which shows that single imputation procedures are generally sub-optimal when dealing with non-random missing data.


When there is not a lot of information about the missing data mechanism, we see the advantage of using FBI over likelihood based approaches (Figure~\ref{fig:2}). The bias and MSE (not shown) for the mean and the variance are smallest for FBI, compared to all other approaches. Recall that in this example, the missing data is only 1\% of the total data. This demonstrates that FBI can better estimate the gene expression and variability needed to identify deferentially expressed genes.

\section{Discussion}
In genomic studies researchers usually have a limited number of samples. They can not sacrifice the information from the data with small number of samples by simply ignoring the observations with missing values, especially when the data is missing not at random. We proposed a fully Bayesian imputation method that works well for small samples sizes. For the purpose of this work, we assume there is one common missing data mechanism for the entire experiment.

In the simulation study we assessed the bias and MSE of the model parameters, performed a sensitivity analysis, and showed that our method is not sensitive to the choice of prior. We recommend using a non-informative uniform prior, but our method can be customized if information about the priors are available. 

We compared this new approach with other existing methods for missing data imputation on an experimental dataset. The real data application revealed that fully Bayesian imputation is better than other methods when there is not a lot of information about the missing data mechanism available. Moreover, it performs similarly to MLE based approaches when the proportion of missing data increases, and hence information about the missing data mechanism increases. In experimental situations, investigators try to limit missing data and often do not have more than 3-4 replicates per condition. Therefore, it is useful to show that this method can work well when there is moderate missingness and is better than other proposed methods when the missingness is limited.

In this work, we assume that each gene has common variance across sample types and gene expressions are normally distributed. It is possible to build a model with a genes by sample types covariance matrix, but this is beyond the scope of this work. The framework of FBI can be extended to non-Normal distributions and also other missing data mechanisms; however, this may require modifications to the model.

\section{Funding}
This work has been partially supported by the National Human Genome Research Institute of the National Institutes of Health under Award Number R00HG006853 (to M.N.M.) and the University of Rochester CTSA award number UL1TR002001 (to M.N.M.) from the National Center for Advancing Translational Sciences of the National Institutes of Health. The content is solely the responsibility of the authors and does not necessarily represent the official views of the National Institutes of Health.

\bibliographystyle{SageV}

\end{document}